\documentclass[twocolumn]{article}

\usepackage[T1]{fontenc}
\usepackage{times}
\usepackage{amsmath,amssymb}
\usepackage{graphicx}
\usepackage{booktabs}
\usepackage[letterpaper,
  textwidth=6.75in,
  textheight=9.25in,
  columnsep=0.25in,
  left=0.75in,
  top=1.0in,
  headheight=16pt,
  headsep=10pt,
  footskip=22pt]{geometry}
\usepackage[hidelinks]{hyperref}
\hypersetup{
  pdftitle={GIA: Germline-Informed Aging with AlphaGenome Finds Genetically Regulated CpGs},
  pdfauthor={Sean Lim}
}

\makeatletter
\renewcommand{\normalsize}{\@setfontsize\normalsize{10pt}{11pt}}
\renewcommand{\small}{\@setfontsize\small{9pt}{10pt}}
\renewcommand{\footnotesize}{\@setfontsize\footnotesize{9pt}{10pt}}
\renewcommand{\large}{\@setfontsize\large{12pt}{14pt}}
\renewcommand{\Large}{\@setfontsize\Large{14pt}{16pt}}
\newcommand{\@runningtitle}{GIA: Germline-Informed Aging with AlphaGenome}
\def\ps@icml{%
  \def\@oddhead{%
    \vbox to \headheight{%
      \vss
      \hbox to \textwidth{\hfil{\small\bfseries \@runningtitle}\hfil}%
      \vskip 3pt
      \hrule height 1pt
    }%
  }%
  \let\@evenhead\@oddhead
  \def\@oddfoot{\hfil\thepage\hfil}%
  \let\@evenfoot\@oddfoot
  \let\@mkboth\@gobbletwo
  \let\sectionmark\@gobble
  \let\subsectionmark\@gobble
}
\renewcommand{\ps@plain}{%
  \def\@oddhead{}%
  \def\@evenhead{}%
  \def\@oddfoot{\hfil\thepage\hfil}%
  \let\@evenfoot\@oddfoot
  \let\@mkboth\@gobbletwo
}
\renewcommand{\@seccntformat}[1]{\csname the#1\endcsname.\hspace{0.35em}}
\renewcommand{\section}{\@startsection{section}{1}{\z@}{-0.12in}{0.02in}{\large\bfseries\raggedright}}
\renewcommand{\subsection}{\@startsection{subsection}{2}{\z@}{-0.10in}{0.01in}{\normalsize\bfseries\raggedright}}
\long\def\@makecaption#1#2{%
  \vskip 10pt
  \noindent{\small{\itshape #1.} #2}\par
}
\renewcommand{\fnum@figure}{Figure~\thefigure}
\renewcommand{\fnum@table}{Table~\thetable}
\makeatother

\normalsize
\newenvironment{icmlabstract}{%
  {\centering\large\bfseries Abstract\par}%
  \vspace{0.06in}%
  \begin{quote}%
}{%
  \par\end{quote}%
  \vspace{0.04in}%
}

\begin{document}
\twocolumn[{%
{\centering
\hrule height 1pt
\vspace{0.18in}
{\Large\bfseries GIA: Germline-Informed Aging with AlphaGenome Finds Genetically Regulated CpGs\par}
\vspace{0.14in}
{\normalsize\bfseries Sean Lim\textsuperscript{1}\par}
\vspace{0.08in}
{\small\textsuperscript{1}Wiess School of Natural Sciences, Rice University, Houston, Texas, USA\par}
\vspace{0.14in}
\hrule height 1pt
\vspace{0.16in}
\par}

\begin{icmlabstract}
Epigenetic clocks estimate age and aging-related phenotypes from DNA methylation at selected CpG sites, but the extent to which these inputs are influenced by germline genetic variation is unclear.
Because methylation at many CpGs is genetically regulated, some between-person variation in clock estimates may reflect inherited genetic differences rather than aging-related change alone.
Here we developed GIA (Germline-Informed Aging), a framework that maps CpGs selected from 13 published epigenetic clocks to blood methylation quantitative trait loci (meQTLs) and scores associated genetic variants with AlphaGenome.
We show that clock CpGs were enriched for blood meQTLs relative to matched unused Illumina 450k probes (62.7\% versus 39.6\%; OR~$2.57$), across multiple clock families, suggesting that age-informative methylation sites are heavily influenced by germline genetic variation.
Ranking by predicted chromatin effect isolated rs10190186, a cis-acting variant at \textit{FHL2} predicted to increase blood chromatin accessibility (ATAC $+1.00$; DNase $+1.64$) and \textit{FHL2} RNA ($+0.30$).
This locus illustrates how inherited variation may shape methylation features repeatedly used by epigenetic clocks, motivating direct tests of whether such variants shift baseline clock estimates or longitudinal aging trajectories.
\end{icmlabstract}
\vspace{-0.08in}
\begin{quote}
\end{quote}
\vspace{0.08in}
}]
\thispagestyle{plain}

\section{Introduction}

Epigenetic clocks use DNA methylation at selected CpG sites to estimate age and aging-related phenotypes.
Models developed to predict chronological age have been complemented by predictors of phenotypic age, mortality risk and the pace of aging, alongside related methylation-based estimators of telomere length~\cite{horvath2013,levine2018,lu2019grimage,belsky2022pace,lu2019telo}.
In many established predictive clocks, CpGs enter the model because they help predict age or an aging-related phenotype, not because genetic regulation of those sites was required for inclusion~\cite{horvath2013,levine2018,belsky2022pace}.
Prediction accuracy leaves open why methylation varies at the selected sites.
That gap matters for interpretation: two people can differ in a clock estimate because methylation at clock sites has changed with age, because inherited genotype set those sites to different baseline methylation values, or because of both.

Large-scale methylation quantitative trait locus (meQTL) studies, including the Genetics of DNA Methylation Consortium (GoDMC), have mapped extensive genetic associations with blood methylation sites~\cite{min2021}.
Longitudinal analyses have shown that many genetic effects on methylation remain stable across the life course~\cite{gaunt2016}.
Genome-wide association studies have also identified loci associated with methylation-based aging biomarkers, some of which share genetic signals with meQTLs at clock CpGs~\cite{mccartney2021}.

What remains unclear is whether clocks use genetically regulated CpGs more often than other probes on the same array, and whether the SNPs at shared clock sites are unusual regulators of chromatin or expression.
Sites used by several clocks are of particular interest, because a genetic association at a shared input could affect more than one biomarker.

AlphaGenome predicts how a DNA variant changes gene expression, chromatin accessibility, transcription-factor binding and related molecular tracks from 1~Mb of sequence context~\cite{avsec2026}.
The recently released AlphaGenome Atlas precomputes those predictions for approximately 9 billion possible single-nucleotide variants, so tissue-specific regulatory effects can be compared among meQTL SNPs without new assays~\cite{cheng2026atlas}.

Here we introduce GIA (Germline-Informed Aging) to test whether published clock CpGs are enriched for blood meQTLs and whether variants at recurrent clock sites have unusual predicted regulatory effects.

\section{Method}

\begin{figure*}[t]
\centering
\includegraphics[width=\textwidth]{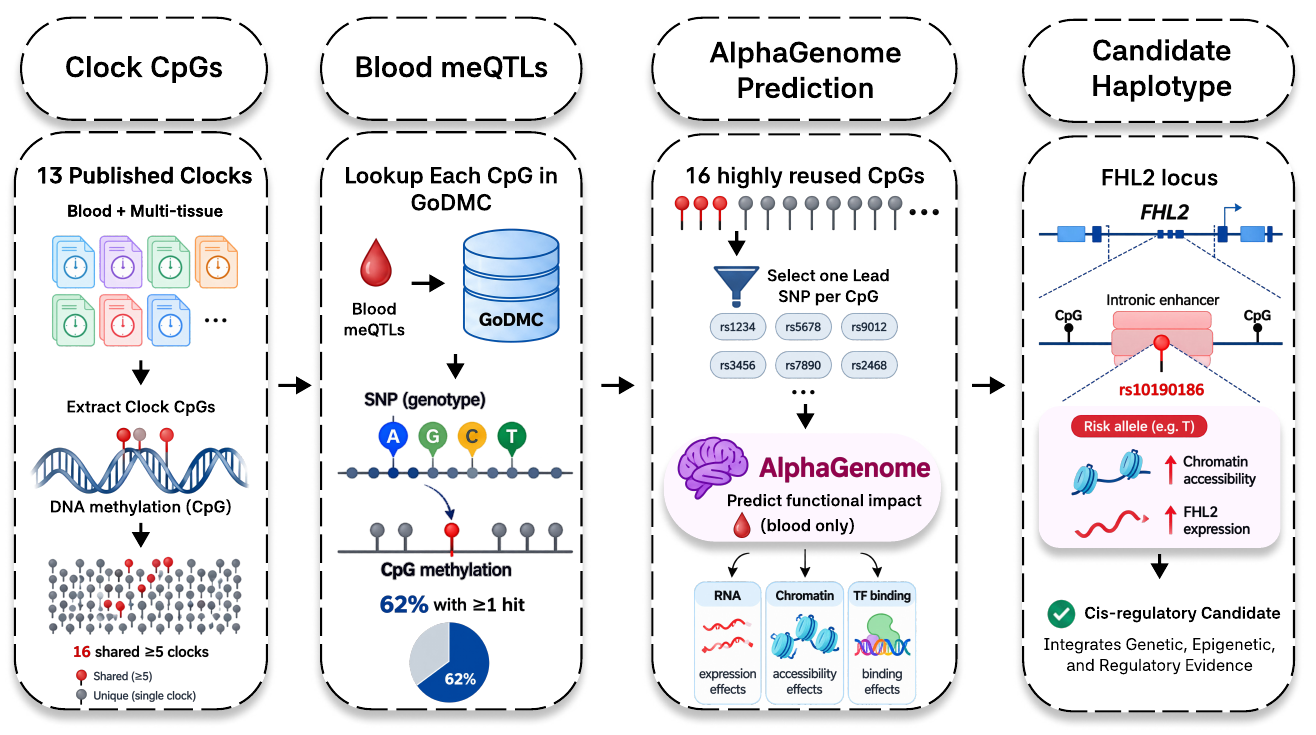}
\caption{GIA pipeline.
CpG sites from 13 published clocks are queried in GoDMC for independent blood meQTLs; lead SNPs at recurrent sites are scored with AlphaGenome.}
\label{fig:pipeline}
\end{figure*}

GIA has three stages.
Published clock coefficients define a set of CpG sites.
Those sites, and a matched set of unused 450k probes, are queried in GoDMC for independent blood meQTLs.
Lead SNPs at sites used by five or more clocks are then scored with AlphaGenome in blood, against meQTL SNPs matched on cis/trans status, minor-allele frequency, and cis distance (Figure~\ref{fig:pipeline}).

\subsection{Problem formulation}

Let $\mathcal{C}$ be the unique clock CpGs present on the Illumina 450k array, and $\mathcal{N}$ a 1:1 matched set of unused 450k probes, $|\mathcal{C}|=|\mathcal{N}|$.
A site is a hit if GoDMC reports at least one independent blood meQTL at the published thresholds.
The primary test is whether clock sites are hits more often than their matches.
Among hits, we compare the lead-SNP effect size $|\beta|$ and the number of independent meQTLs.
At the SNP layer, the 15 lead SNPs at CpGs used by $\geq$5 clocks are compared with 15 matched blood-meQTL SNPs on AlphaGenome blood summaries.

\subsection{Clock CpGs}

CpG identifiers were taken from Biolearn coefficient tables~\cite{ying2025biolearn} for 13 published blood or multi-tissue clocks (Supplementary Table~\ref{tab:clocks}): chronological (Horvath 2013, Horvath 2018 skin and blood, Hannum, Lin, Vidal-Bralo, GP-age 71); phenotypic (Levine PhenoAge, HRSInCH PhenoAge); mortality (GrimAge V1); pace (DunedinPACE, DunedinPoAm38); telomere (DNAmTL); and stochastic (StocZ).
We omitted mitotic clocks (EpiTOC, MiAge); Ying CausAge, DamAge, and AdaptAge, which select meQTLs by design; GrimAge V2, Stochastic P, and Stochastic H, whose CpG sets duplicate V1, PhenoAge, and Horvath 2013; Zhang's 10-CpG clock; the nested GP-age 10, 30, A, B, and C models, which are subsets of GP-age 71; and Weidner, Garagnani, and Bocklandt (1--3 CpGs, all already present in larger clocks).
Unique cg identifiers were retained; a site used by $k$ clocks has sharing $n_{\mathrm{clocks}}=k$.

\subsection{Blood meQTLs}

Clock CpGs were queried in GoDMC~\cite{min2021} for independent (clumped) whole-blood associations, using the published cis ($\textit{P} < 1 \times 10^{-8}$) and trans ($\textit{P} < 1 \times 10^{-14}$) thresholds.
GoDMC reports the change in residual methylation, in standard-deviation units, per copy of the effect allele.
Cis associations are those within 1~Mb of the CpG.
The lead SNP at each hit CpG is the independent association with smallest \textit{P}, then largest $|\beta|$.
Insertion/deletion alleles were excluded.
A missing GoDMC row is not evidence against genetic regulation: the catalog is 450k-based and omits associations weaker than these cutoffs.

\subsection{Matched null}

The null universe is the Illumina HumanMethylation450 BeadChip~\cite{bibikova2011} (hg19), restricted to autosomes.
Clock probes were removed.
Each clock CpG on the array was paired, without replacement, to an unused probe on the same chromosome and CpG-island class (island, shore $\leq$2~kb, shelf 2--4~kb, open sea), using a fixed random seed.
If a chromosome--island stratum had too few unused probes, matching fell back to chromosome only.
Matched null CpGs were queried in GoDMC at the same thresholds.
For AlphaGenome, each scored clock lead was paired 1:1 with a null-CpG lead of the same cis/trans class, excluding the same rsID, by minimizing a cost of minor-allele-frequency difference (allowed gap $\leq 0.08$) and, for cis SNPs, the squared log ratio of SNP--CpG distances.

\subsection{Statistical tests}

Hit rate and cis versus trans among associations were tested with two-sided Fisher's exact tests.
Lead $|\beta|$, the number of independent meQTLs per hit CpG, and AlphaGenome summaries were compared with two-sided Mann--Whitney tests.
Cliff's $\Delta = P(X>Y)-P(X<Y)$ reports stochastic dominance of clock values $X$ over null values $Y$~\cite{cliff1993}.
Each clock was tested against that clock's own matched nulls; Benjamini--Hochberg $q$ was taken across the 13 clocks~\cite{benjamini1995}.
AlphaGenome $q$ was taken across the ten blood summaries only.
CpG-level and SNP-level tests were not pooled into one FDR.

\subsection{AlphaGenome scoring}

GoDMC coordinates are hg19.
Lead SNPs were lifted to hg38 with Ensembl~\cite{yates2026} and scored REF versus ALT using AlphaGenome~\cite{avsec2026} with the recommended RNA-seq, DNase, ATAC, and TF ChIP scorers.
Tracks were restricted to blood by ontology and biosample name.
The quantile score is the percentile of the predicted effect among background variants, bounded at 1.
For each SNP we summarized median and maximum $|\mathrm{quantile}|$ within each assay, and the median and 90th percentile of $|\mathrm{quantile}|$ across blood tracks.
Because the strongest tracks saturate (max RNA-seq $|\mathrm{quantile}| \geq 0.95$ at all 15 scored 5+ SNPs), recurrent sites were ordered by $n_{\mathrm{clocks}}$ and then by the maximum absolute raw chromatin score among blood ATAC, DNase, and TF ChIP.
The top-ranked variant was annotated for gene, enhancer overlap, and other clock CpGs sharing the same lead SNP.

\section{Results}

\subsection{Most clock CpGs overlap blood meQTLs despite limited reuse across clocks}

We compiled the CpG sites used by 13 published blood or multi-tissue epigenetic clocks, yielding 3,837 unique sites from 4,366 clock--site assignments in total.
Despite measuring related aging phenotypes, the clocks largely relied on distinct CpGs. 
Of the 3,837 unique sites, 3,488 were private to a single clock, 239 were shared by two clocks, and only 16 appeared in five or more. 
The most widely reused site, cg09809672, was included in nine clocks, followed by cg19722847 in seven. 

\begin{figure*}[t]
\centering
\includegraphics[width=\textwidth,height=0.44\textheight,keepaspectratio]{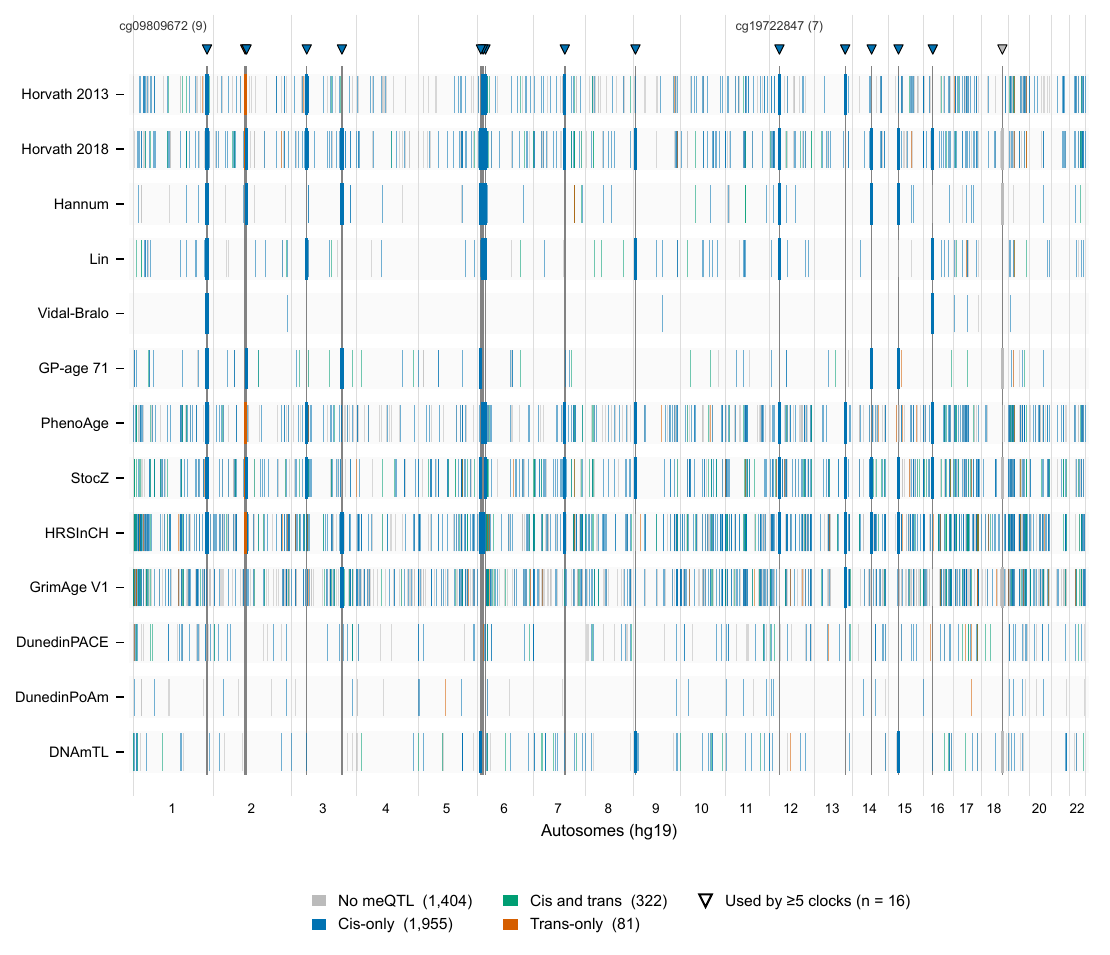}
\caption{Autosomal clock CpGs on the Illumina 450k array (hg19; $n = 3{,}762$), colored by GoDMC class.
This is the subset used for the matched analysis, not the full 3,837 unique clock CpGs.
Triangles mark the 16 sites used by five or more clocks.}
\label{fig:map}
\end{figure*}

To determine how frequently these clock CpGs are under detectable germline regulation, we queried the Genetics of DNA Methylation Consortium (GoDMC) for independent blood methylation quantitative trait loci (meQTLs), using the published cis ($\textit{P} < 1 \times 10^{-8}$) and trans ($\textit{P} < 1 \times 10^{-14}$) thresholds.
Of the 3,837 unique CpGs, 2,365 (62\%) had at least one independent blood meQTL.
Most were regulated only by a nearby variant (1,961 cis-only), whereas 82 had only trans associations and 322 had both cis and trans signals.
Figure~\ref{fig:map} maps the 3,762 autosomal 450k CpGs subsequently used for the matched analysis.

\subsection{Epigenetic clocks preferentially use genetically regulated methylation sites}

Because many 450k probes have meQTLs, we compared 3,762 clock CpGs on the array 1:1 with unused probes on the same chromosome and CpG-island class, using the same GoDMC thresholds.

Clock CpGs were substantially more likely to have a blood meQTL than their matched probes (2,358/3,762, 62.7\%, versus 1,488/3,762, 39.6\%; OR 2.57, Fisher's $\textit{P} = 3.8 \times 10^{-90}$; Table~\ref{tab:tests}).
The enrichment was particularly pronounced among the 16 CpGs reused by five or more clocks: 15 of 16 had a blood meQTL compared with 4 of 16 matched probes (OR 45.0, $\textit{P} = 1.7 \times 10^{-4}$).
Thus, CpGs retained by published epigenetic clocks are disproportionately enriched for detectable germline regulation, with recurrent clock sites showing especially strong enrichment.

\begin{table}[t]
\centering
\caption{Clock versus matched-null tests.
CpG-level tests use 450k-matched sites.
AlphaGenome tests are 15 versus 15 SNPs; $q$ is across those ten summaries only.}
\label{tab:tests}
\scriptsize
\setlength{\tabcolsep}{2.2pt}
\resizebox{\columnwidth}{!}{%
\begin{tabular}{@{}l c c c c c@{}}
\toprule
Test & Clock & Null & Effect & \textit{P} & $q$ \\
\midrule
\multicolumn{6}{@{}l}{\textit{GoDMC}} \\
Hit rate, all & 2,358/3,762 & 1,488/3,762 & OR 2.57 & $3.8 \times 10^{-90}$ & --- \\
Hit rate, $\geq$5 clocks & 15/16 & 4/16 & OR 45.0 & $1.7 \times 10^{-4}$ & --- \\
Lead $|\beta|$ & 0.247 & 0.203 & $\Delta$ 0.13 & $1.7 \times 10^{-11}$ & --- \\
Indep.\ meQTLs & 1 & 1 & $\Delta$ 0.17 & $4.7 \times 10^{-24}$ & --- \\
Cis among hits & 3,436/3,978 & 1,919/2,105 & OR 0.61 & $2.5 \times 10^{-8}$ & --- \\
\midrule
\multicolumn{6}{@{}l}{\textit{AlphaGenome ($n = 15$)}} \\
Median $|\mathrm{q}|$, all & 0.544 & 0.482 & $\Delta$ 0.38 & 0.081 & 0.27 \\
90th pct.\ $|\mathrm{q}|$ & 0.912 & 0.841 & $\Delta$ 0.39 & 0.071 & 0.27 \\
RNA-seq, median & 0.567 & 0.442 & $\Delta$ 0.44 & 0.040 & 0.27 \\
RNA-seq, max & 0.995 & 0.995 & $\Delta$ 0.06 & 0.80 & 0.89 \\
DNase, median & 0.364 & 0.407 & $\Delta$ 0.09 & 0.69 & 0.87 \\
DNase, max & 0.921 & 0.863 & $\Delta$ 0.10 & 0.65 & 0.87 \\
ATAC, median & 0.670 & 0.652 & $\Delta$ 0.01 & 0.97 & 0.97 \\
ATAC, max & 0.897 & 0.824 & $\Delta$ 0.16 & 0.47 & 0.87 \\
TF ChIP, median & 0.567 & 0.421 & $\Delta$ 0.12 & 0.59 & 0.87 \\
TF ChIP, max & 0.900 & 0.894 & $\Delta$ 0.17 & 0.44 & 0.87 \\
\bottomrule
\end{tabular}}
\end{table}

Among CpGs with a meQTL, the strongest independent SNP at clock sites showed a larger absolute effect on methylation than that at matched probes (median $|\beta| = 0.247$ versus 0.203; Mann--Whitney $\textit{P} = 1.7 \times 10^{-11}$).
Clock CpGs also tended to have more independent meQTL associations (Cliff's $\Delta = 0.17$, $\textit{P} = 4.7 \times 10^{-24}$; Table~\ref{tab:tests}).
In contrast, their associations were somewhat less restricted to cis regulation: 86\% of clock-site associations were cis compared with 91\% among matched-null associations (OR 0.61, $\textit{P} = 2.5 \times 10^{-8}$).
These results indicate that the difference between clock and background CpGs extends beyond meQTL presence alone.

\begin{figure}[t]
\centering
\includegraphics[width=\columnwidth]{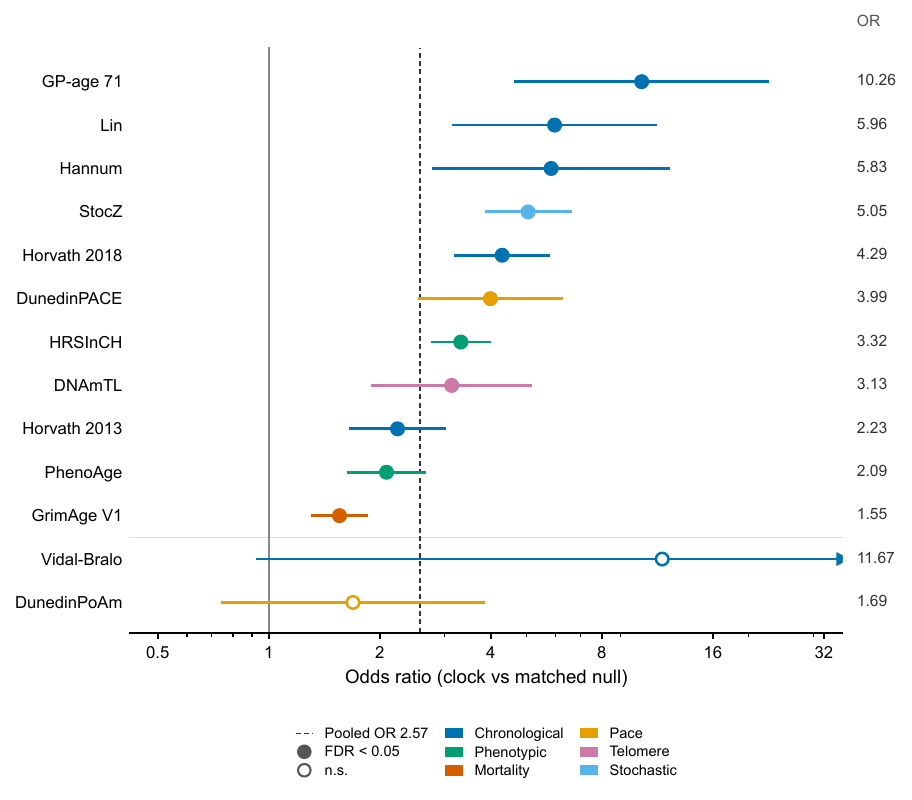}
\caption{Per-clock meQTL enrichment versus each clock's matched nulls.
Filled points, FDR $< 0.05$ across 13 clocks; open, not significant.
Dashed line, pooled odds ratio $2.57$.}
\label{fig:forest}
\end{figure}

To determine whether the pooled enrichment was driven by one clock or one class of clocks, we repeated the analysis separately for each model using its own matched-null set (Figure~\ref{fig:forest}).
The enrichment was observed across chronological, phenotypic, pace-of-aging, telomere and stochastic clock families.
Eleven of the 13 clocks remained significant after FDR correction, indicating that the overall signal was not driven by a single large clock.
GrimAge V1, despite containing the largest unique CpG set, showed a comparatively modest enrichment (OR 1.55), whereas several other clocks showed substantially larger effects.
Together, these analyses suggest that preferential use of genetically regulated CpGs is a recurring feature across otherwise distinct epigenetic clock designs.

\subsection{Clock-meQTL SNPs show no class-wide regulatory enrichment}

For the 16 CpGs reused by five or more clocks, we took the strongest associated SNP at each site (smallest meQTL $\textit{P}$, then largest $|\beta|$).
Fifteen of the 16 had a GoDMC meQTL.
Their lead SNPs were scored with AlphaGenome across blood RNA-seq, DNase, ATAC and transcription-factor ChIP tracks and compared with blood-meQTL SNPs matched on cis/trans status, minor-allele frequency and, for cis variants, distance to the CpG.

No AlphaGenome summary distinguished the two groups after FDR correction (Table~\ref{tab:tests}).
The largest nominal difference was the median RNA-seq score ($\textit{P} = 0.040$), which did not survive correction ($q = 0.27$); all other unadjusted $\textit{P}$ values were at least 0.071.

\subsection{A cis-acting \textit{FHL2} variant stands out among recurrent clock-site meQTLs}

\begin{figure*}[t]
\centering
\includegraphics[width=0.72\textwidth]{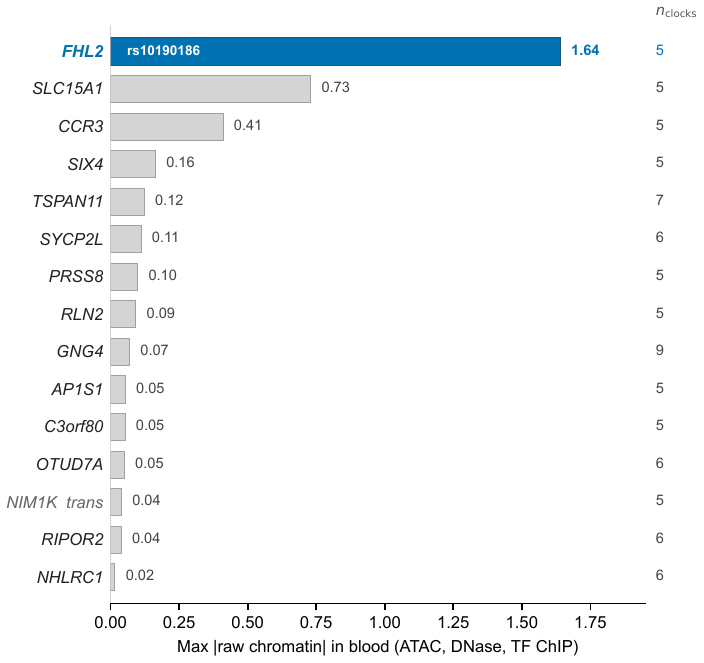}
\caption{Lead SNPs at CpGs used by five or more clocks, ranked by maximum absolute AlphaGenome raw chromatin in blood (ATAC, DNase, TF ChIP).
\textit{FHL2} (rs10190186) is highlighted.
\textit{NIM1K} is the sole trans association; cg19283806 has no GoDMC row and is omitted.}
\label{fig:rank}
\end{figure*}

AlphaGenome quantile scores gave little resolution among the 15 recurrent-site meQTL SNPs because the strongest blood tracks were already near saturation; maximum RNA-seq $|\mathrm{quantile}|$ was at least 0.95 for all 15 variants.
Ranking by raw predicted chromatin effect isolated rs10190186 at \textit{FHL2} (Figure~\ref{fig:rank}).

rs10190186 is an A$>$G intronic variant in \textit{FHL2} that overlaps the Ensembl enhancer ENSR2\_BQWN4 (chr2:105,380,928, hg38)~\cite{yates2026}.
It is the strongest blood meQTL for cg06639320, a CpG included in five epigenetic clocks, and had the largest AlphaGenome chromatin prediction among the recurrent-site lead SNPs (DNase raw score $+1.64$; Figure~\ref{fig:rank} and Table~\ref{tab:core}).
The same allele is also the strongest meQTL for cg22454769, located only 28~bp away and used by four clocks.
Both CpGs lie within \textit{FHL2}, approximately 18~kb from rs10190186.

\begin{table}[ht]
\centering
\caption{Sites used by five or more clocks, ranked by sharing.
Lead SNP: smallest \textit{P}, then largest $|\beta|$.
$|\mathrm{chr.}|$ is max absolute AlphaGenome raw chromatin in blood.
cg19283806 has no GoDMC row; cg22809047 is trans (all other leads are cis).}
\label{tab:core}
\scriptsize
\setlength{\tabcolsep}{2pt}
\resizebox{\columnwidth}{!}{%
\begin{tabular}{@{}l c l r r l@{}}
\toprule
CpG & $n$ & SNP & $\beta$ & $|\mathrm{chr.}|$ & Gene \\
\midrule
cg09809672 & 9 & rs114694761 & $+0.19$ & 0.07 & \textit{GNG4} \\
cg19722847 & 7 & rs11050984 & $+0.11$ & 0.12 & \textit{TSPAN11} \\
cg04875128 & 6 & rs12101830 & $-0.14$ & 0.05 & \textit{OTUD7A} \\
cg06493994 & 6 & rs4712938 & $-0.15$ & 0.04 & \textit{RIPOR2} \\
cg16867657 & 6 & rs6920155 & $+0.14$ & 0.11 & \textit{SYCP2L} \\
cg19283806 & 6 & --- & --- & --- & --- \\
cg22736354 & 6 & rs189617077 & $-0.36$ & 0.02 & \textit{NHLRC1} \\
cg03032497 & 5 & rs4901980 & $-0.14$ & 0.16 & \textit{SIX4} \\
cg04084157 & 5 & rs1048303 & $+0.09$ & 0.05 & \textit{AP1S1} \\
cg04836038 & 5 & rs9513513 & $-0.07$ & 0.73 & \textit{SLC15A1} \\
cg06639320 & 5 & rs10190186 & $+0.33$ & 1.64 & \textit{FHL2} \\
cg07211259 & 5 & rs10758690 & $-0.41$ & 0.09 & \textit{RLN2} \\
cg07553761 & 5 & rs147828440 & $-0.44$ & 0.05 & \textit{C3orf80} \\
cg10917602 & 5 & rs112750937 & $-0.26$ & 0.10 & \textit{PRSS8} \\
cg22809047 & 5 & rs10058350 & $+0.17$ & 0.04 & \textit{NIM1K} \\
cg26614073 & 5 & rs114933663 & $-1.20$ & 0.41 & \textit{CCR3} \\
\bottomrule
\end{tabular}}
\end{table}

Notably, the G allele increased residual methylation at both CpGs ($+0.33$ and $+0.53$ SD), in the same direction as their previously reported age-associated hypermethylation~\cite{kananen2016,marttila2015}.
AlphaGenome also predicted concordant regulatory effects in blood, including increased chromatin accessibility (ATAC $+1.00$ in CD4 T cells; DNase $+1.64$ in CD34$^{+}$ myeloid progenitors) and increased \textit{FHL2} RNA expression ($+0.30$).
rs10190186 is therefore a candidate cis-regulatory variant at two recurrent clock CpGs. Whether it shifts any clock's age estimate was not tested.

\section{Discussion}

Clock CpGs were enriched for blood meQTLs relative to matched unused array probes (OR~$2.57$), including across clock families and among sites reused by several models.
Lead SNPs at those recurrent sites were indistinguishable from other blood meQTLs in the AlphaGenome summaries we tested.
Ranking within the recurrent set pointed to rs10190186 at \textit{FHL2}.

The comparison here is: whether the CpGs inside published clocks are enriched for meQTLs relative to unused probes.
Ying and colleagues found that existing clocks are not enriched for CpGs with putative causal effects on aging-related traits~\cite{ying2024}.
A clock CpG can therefore be genetically regulated without its methylation driving aging.

Shared sites are useful because one association could enter several clock outputs.
They are not independent replications: the models share CpGs and the same array.
If many clock CpGs have stable meQTLs~\cite{gaunt2016}, between-person differences in clock estimates could include inherited baseline methylation as well as aging-related change.
That split has to be measured in people with genotype and methylation, using each clock's published weights and preprocessing~\cite{horvath2013}; the fraction of CpGs with meQTLs is not the fraction of clock variance explained by genotype.

AlphaGenome scored sequence effects on blood chromatin and expression, not methylation or aging~\cite{avsec2026}.
The 15-versus-15 comparison is small, and both groups were already meQTLs.
Its use here is to rank hypotheses among recurrent-site leads.

Hypermethylation of cg06639320 and cg22454769 with age has previously been reported in blood~\cite{kananen2016}, and islet work has linked \textit{FHL2} methylation to expression~\cite{bacos2016}.
rs10190186 is the shared lead meQTL at both nearby clock CpGs; the G allele is associated with higher methylation, and AlphaGenome predicts increased blood accessibility and \textit{FHL2} RNA.
That is a reason to test the locus, not a claim that we discovered \textit{FHL2} aging or that the allele accelerates organismal aging.
The SNP may tag another variant, enhancer overlap is not a functional proof, and the predictions mix blood cell types.
Fine-mapping, colocalization, and perturbation remain to be done.

The enrichment analysis used autosomal 450k probes and thresholded whole-blood meQTLs, matched on chromosome and island class.
Unused probes are not an age-associated control, the recurrent subset is small, and other tissues and populations were not tested.

Whether rs10190186, or other clock-site meQTLs, associate with baseline clock estimates, longitudinal change, or neither can be asked in genotype--methylation cohorts while keeping each clock's published scoring.

\section{Conclusion}

Published clocks select CpGs because they track age.
Those inputs are enriched for blood meQTLs relative to matched unused 450k probes, including across clock families; sites used by five or more models are almost all meQTLs.
Their lead SNPs look like ordinary blood meQTLs in AlphaGenome.
rs10190186 at \textit{FHL2} ranked highest on predicted chromatin effect and is associated with higher methylation at two nearby clock CpGs.
Whether that allele, or others like it, shifts a clock's age number remains to be measured.

\section*{Acknowledgements}

The author acknowledges the Genetics of DNA Methylation Consortium, the Biolearn maintainers, and the AlphaGenome developers for publicly available resources used in this work.
This study was conducted at Rice University.

\section*{Funding}

This work received no specific grant from any funding agency in the public, commercial, or not-for-profit sectors.

\section*{Conflict of interest}

The author declares no competing interests.

\section*{Data availability}

Clock CpG sets were taken from published Biolearn coefficient tables~\cite{ying2025biolearn}.
Independent blood meQTLs were obtained from the Genetics of DNA Methylation Consortium~\cite{min2021}.
Variant-effect predictions were generated with AlphaGenome~\cite{avsec2026}.
Derived tables supporting the figures and numbered tables, including clock--CpG mappings, GoDMC lookups, matched-null assignments, enrichment tests, and AlphaGenome summaries, are provided in the project repository.

\clearpage
\onecolumn
\section*{Supplementary information}
\setcounter{table}{0}
\refstepcounter{table}
\renewcommand{\thetable}{S\arabic{table}}
\noindent{\small\textit{Supplementary Table~\thetable.}
Published clocks included in GIA.
CpG counts are unique cg identifiers from Biolearn~\cite{ying2025biolearn}.}\par
\label{tab:clocks}
\vspace{8pt}
{\centering
\footnotesize
\setlength{\tabcolsep}{4pt}
\begin{tabular}{@{}l l l r l@{}}
\toprule
Clock & Target & Tissue & CpGs & Reference \\
\midrule
Horvath 2013 & Chronological age & Multi-tissue & 353 & \cite{horvath2013} \\
Horvath 2018 & Chronological age & Skin and blood & 391 & \cite{horvath2018} \\
Hannum & Chronological age & Blood & 69 & \cite{hannum2013} \\
Lin & Chronological age & Blood & 99 & \cite{lin2016} \\
Vidal-Bralo & Chronological age & Blood & 8 & \cite{vidalbralo2016} \\
GP-age 71 & Chronological age & Blood & 71 & \cite{varshavsky2023} \\
Levine PhenoAge & Phenotypic age & Blood & 513 & \cite{levine2018} \\
HRSInCH PhenoAge & Phenotypic age & Blood & 959 & \cite{higginschen2022} \\
GrimAge V1 & Mortality & Blood & 1,030 & \cite{lu2019grimage} \\
DunedinPACE & Pace of aging & Blood & 173 & \cite{belsky2022pace} \\
DunedinPoAm38 & Pace of aging & Blood & 46 & \cite{belsky2020poam} \\
DNAmTL & Telomere length & Blood & 140 & \cite{lu2019telo} \\
StocZ & Stochastic aging & Blood & 514 & \cite{tong2024} \\
\bottomrule
\end{tabular}\par
}

\end{document}